\documentclass[3p,times]{elsarticle}

\usepackage{ecrc}

\volume{00}

\firstpage{1}

\journalname{Nuclear Physics A}

\runauth{}

\jid{npa}

\jnltitlelogo{Nuclear Physics A}

\usepackage{amssymb}

\biboptions{square,comma,numbers,sort&compress}

\usepackage[figuresright]{rotating}

\usepackage{graphicx}
\usepackage{dcolumn}
\usepackage{bm}
\usepackage{amssymb}
\usepackage{xspace}

\usepackage{hyperref}
\usepackage{amsmath}

\usepackage{color}

\newcommand{\pp}{\ensuremath{\rm pp}\xspace}

\newcommand{\ppb}{\ensuremath{\rm p\!-\!Pb}\xspace}
\newcommand{\pbpb}{\ensuremath{\rm Pb\!-\!Pb}\xspace}

\newcommand{\pt}{\ensuremath{p_{\rm{T}}}\xspace}

\newcommand{\dedx}{\ensuremath{{\rm d}E/{\rm d}x}\xspace}

\newcommand{\ptopi}{\ensuremath{{\rm p } / \pi}\xspace}

\begin{document}

\begin{frontmatter}



\dochead{}

\title{Production of $\pi/{\rm K/p}$ from intermediate to high \pt in pp, p--Pb and Pb--Pb collisions measured by ALICE}
\author[label1]{Antonio Ortiz Velasquez (for the ALICE Collaboration)}
\ead{aortizve@cern.ch}
\address{Lund University, Department of Physics, Division of Particle Physics Box 118, SE-221 00, Lund, Sweden}
\address[label1]{Now at:  Instituto de Ciencias Nucleares, Universidad Nacional Aut\'onoma de M\'exico,  Apartado Postal 70-543, M\'exico
D. F. 04510, M\'exico.}




\begin{abstract}

In this work the results on the transverse momentum distributions  ($\sim0.3<p_{\rm T}<15$ GeV/$c$) of charged pions, kaons and (anti)protons,  measured in $\rm pp$ and \pbpb collisions at $\sqrt{s_{\rm NN}}=2.76$ TeV and in \ppb collisions at $\sqrt{s_{\rm NN}}=5.02$ TeV, are presented. The evolution of the spectral shapes and particle ratios with multiplicity is shown and the similarities among the different systems are discussed.

\end{abstract}


\end{frontmatter}




\section{Introduction}
\label{sec:1}

\begin{figure}[htbp]
\begin{center}
   \includegraphics[width=0.33\textwidth]{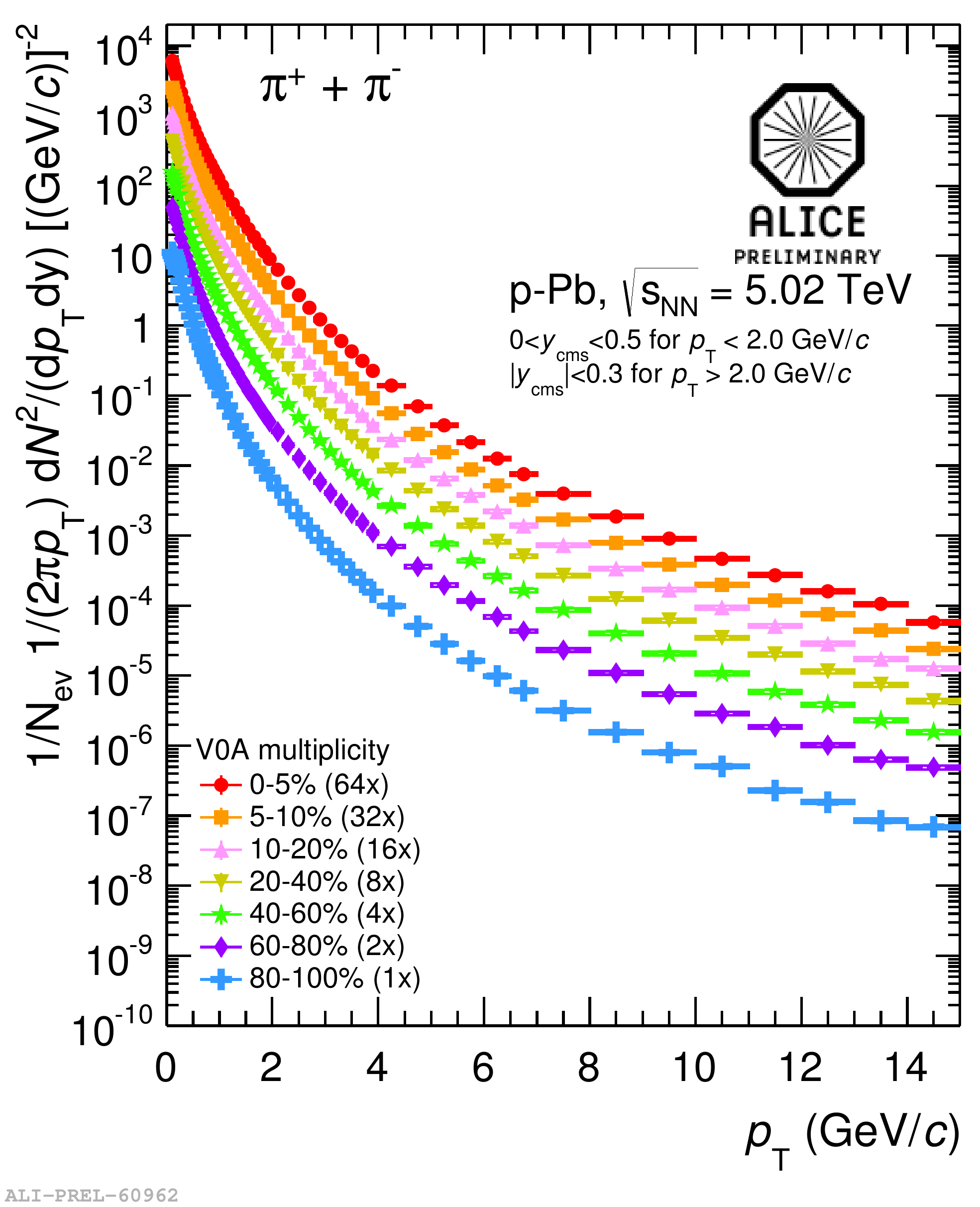}
   \includegraphics[width=0.33\textwidth]{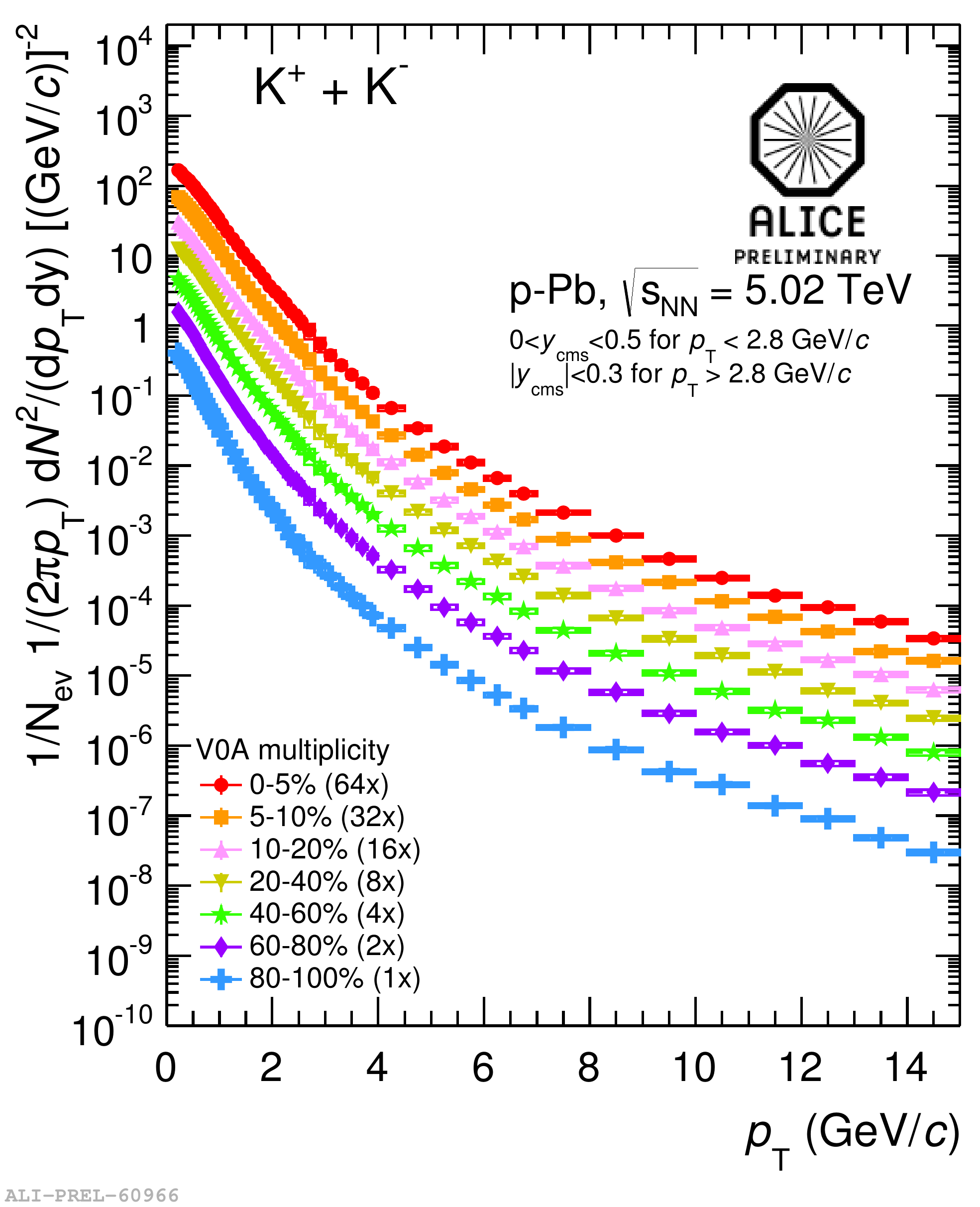}   
   \includegraphics[width=0.33\textwidth]{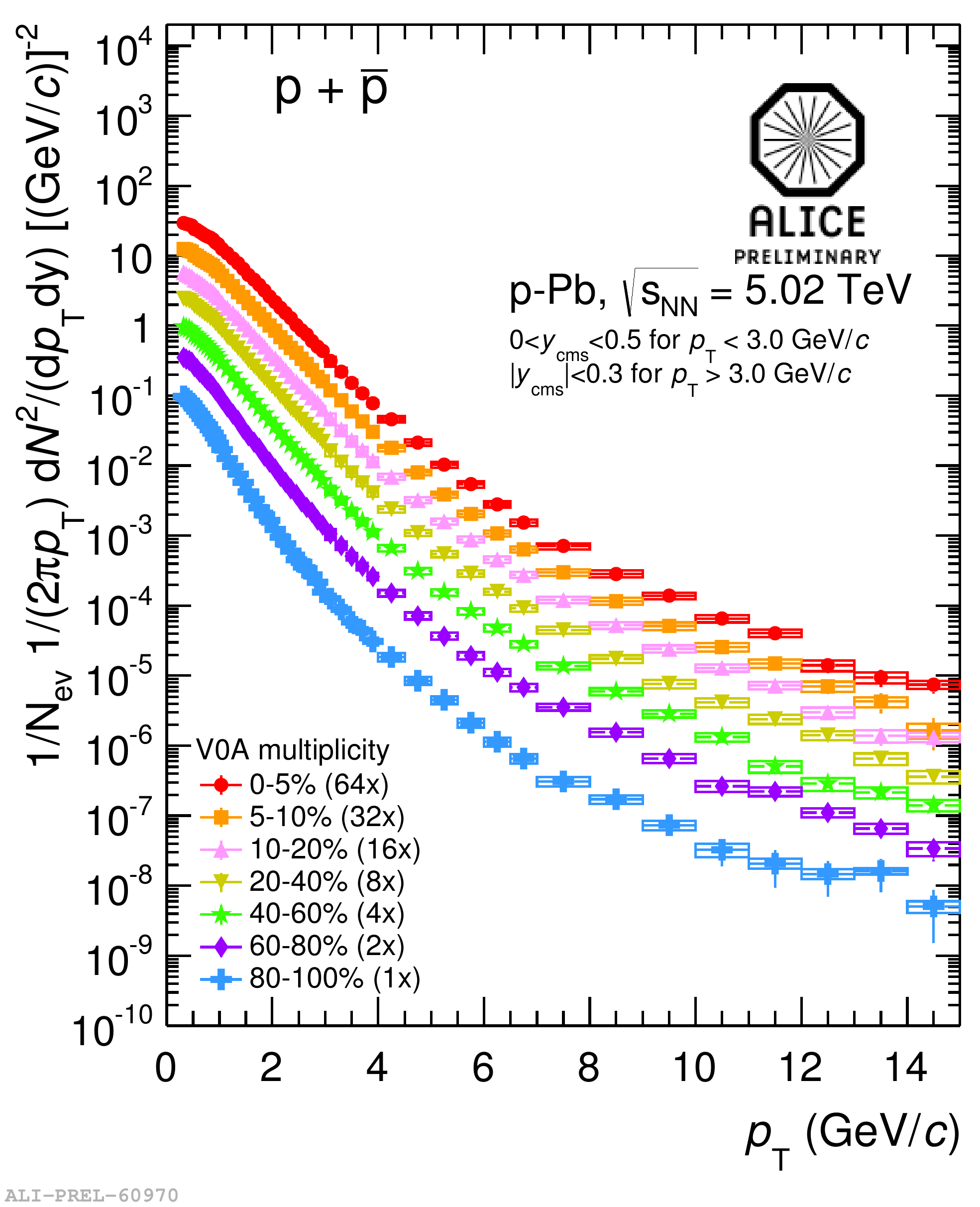}   
   \caption{\pt spectra of charged pions (left), kaons (middle) and (anti)protons (right) measured in \ppb collisions at $\sqrt{s_{\rm NN}}$ = 5.02 TeV for different V0A multiplicity event classes. The systematic and statistical error are plotted as color boxes and vertical error bars, respectively.}
  \label{Fig1}
\end{center}
\end{figure}

In central heavy ion collisions at ultra relativistic energies it is well established that a strongly interacting medium of quarks and gluons is created. The transverse momentum, \pt, distributions of identified hadrons contain valuable information about the collective expansion of the system ($\pt\lesssim2$ GeV/$c$), the presence of new hadronization mechanisms like quark recombination ($2\lesssim \, \pt \lesssim \,8$ GeV/$c$)~\cite{Fries:2008hs} and, at larger transverse momenta, the possible modification of the fragmentation due to the medium~\cite{Sapeta:2007ad,Bellwied:2010pr}. ALICE has reported the  transverse momentum spectra, as a function of the collision centrality, of charged pions, kaons and (anti)protons from low (hundreds of MeV/$c$)~\cite{Abelev:2013vea} to high (20 GeV/$c$)~\cite{Abelev:2014laa} \pt. From the analysis of the low \pt results in the most central collisions, the radial flow, $\langle \beta_{\rm T} \rangle$, is found to be $\approx10$\% higher than at RHIC,  while the kinetic freeze-out temperature was found to be comparable to that extracted from data at RHIC, $T_{\rm kin}=$  95 MeV~\cite{Abelev:2013vea}.  The spectra are well described by hydrodynamic models, except the low \pt ($<1$ GeV/$c$) proton yield~\cite{Bozek:2012qs,Karpenko:2012yf,Werner:2012xh,Shen:2011eg}. Models which best describe the data include hadronic rescattering with non-negligible antibaryon annihilation~\cite{Werner:2012xh,Shen:2011eg}. For intermediate to high \pt ($>3$ GeV/$c$), the spectra develop the power law tail  characteristic of hard partonic processes. The proton-to-pion ratio increases from $\approx 0.38$ to $\approx 0.8$ going from peripheral (60-80\%) to central (0-5\%) \pbpb collisions at $\pt \approx 3$ GeV/$c$, then decreases to the value measured for vacuum fragmentation (\pp collisions) for $\pt > 10$ GeV/$c$. The result obtained for the most central collisions is similar to that measured at RHIC~\cite{Adare:2013esx,Abelev:2006jr}. The kaon-to-pion ratio also exhibits a small bump around $\pt=$ 3 GeV/$c$. Quark recombination does not predict the latter effect and fails to describe the \ptopi ratio over the entire \pt range.

Surprisingly, early results from LHC showed that \ppb collisions exhibit behaviours reminiscent to those due to final state effects, namely, hints of collective effects (radial and elliptic flow, ridge structure), but no sign of jet quenching ~\cite{ABELEV:2013wsa,Abelev:2013haa}. Hence, the main focus of this work is to present complementary measurements on identified particle production, $\pi/{\rm K/p}$, from intermediate \pt, 2-3 GeV/$c$, to high \pt,  $<$ 15 GeV/$c$, in \ppb collisions at $\sqrt{s}_{\rm NN}$ = 5.02 TeV.
%
%
%

\section{Analysis method}

\begin{figure}[htbp]
\begin{center}
   \includegraphics[width=0.72\textwidth]{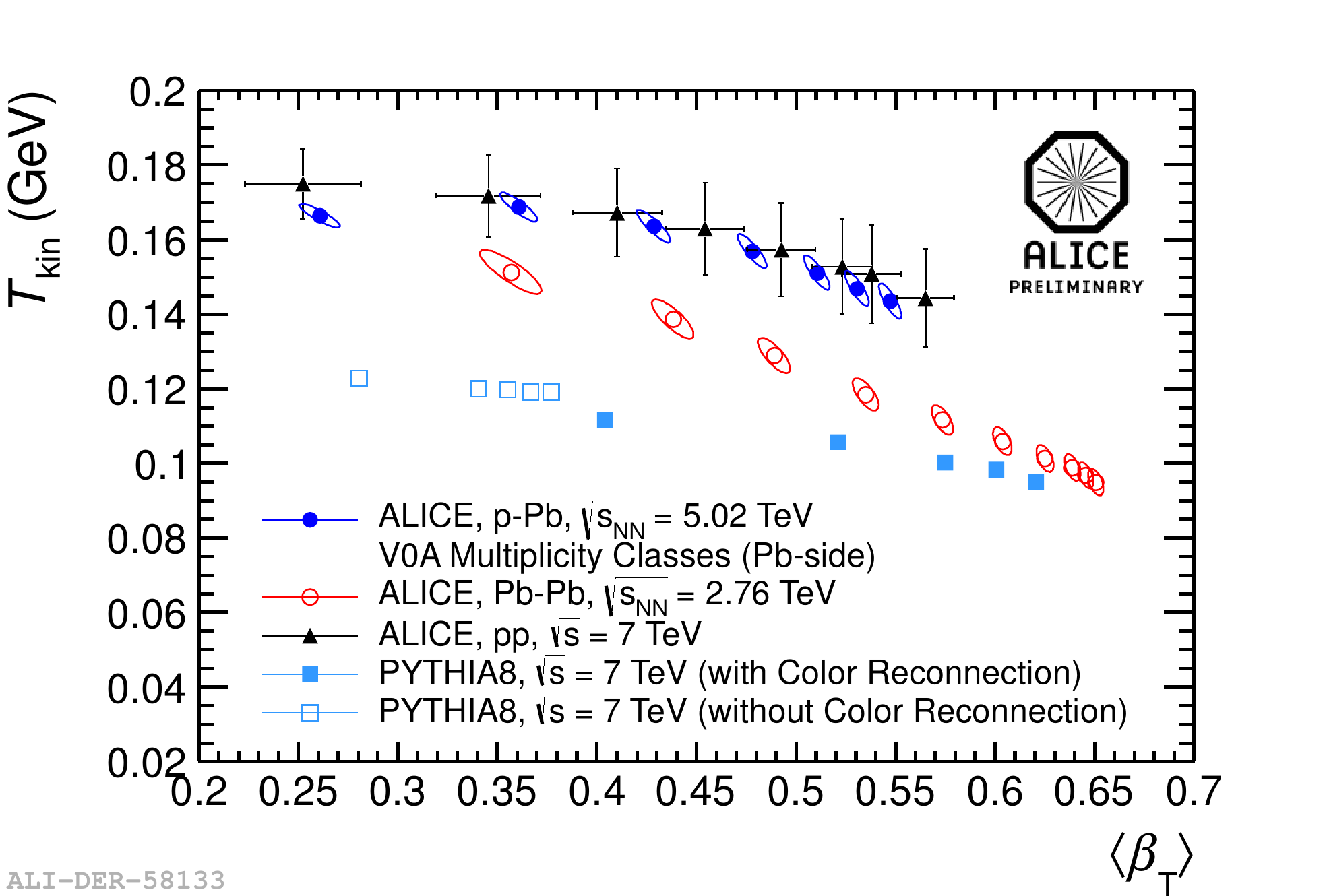}
   \caption{Comparison of the results from the blast-wave analysis applied to all available systems: \pp, \ppb and \pbpb collisions. The spectral shape analysis was also applied to Pythia 8 events. Charged-particle multiplicity increases from left to right}
  \label{Fig2}
\end{center}
\end{figure}

Data from \ppb collisions were obtained using the ALICE detector~\cite{Aamodt:2008zz}.  In the analysis presented here, we have used 82 millions minimum bias (MB) events. The MB trigger signal was provided by the VZERO counters, two arrays of 32 scintillator tiles each covering the full azimuth within $2.8 < \eta_{\rm lab} < 5.1$ (V0A, Pb beam direction) and $-3.7 < \eta_{\rm lab} < -1.7$ (V0C, p beam
direction). The signal amplitude and arrival time collected in each tile were recorded. A coincidence of signals in both V0A and V0C detectors was required to remove contamination from single diffractive and electromagnetic events.
For \ppb collisions, the nucleon-nucleon center of mass system has a rapidity of $y_{\rm NN}=-0.465$ in the direction of the proton beam. Due to the weak correlation between geometry and multiplicity, the particle production is studied in seven multiplicity event classes instead of centrality, this selection was based on cuts on the total charge deposited in the V0A detector~\cite{Abelev:2013haa}. Particle identification at high \pt is done using the specific energy loss, \dedx, measured in the Time Projection Chamber (TPC) for tracks in $|y_{\rm cms}|<0.3$. The yields (${\rm d}N/{\rm d}y{\rm d}\pt$), particle ratios as a function of \pt and their uncertainties  are obtained using the method discussed in this reference~\cite{Abelev:2014laa}. The main quantities to determine the yields are the particle fractions as a function of \pt, {\it i.e.,} the contribution of a given particle species to the yield of inclusive charged particles in a given \pt interval. They are determined from a four-Gaussian (pions, kaons, protons and electrons) fits to the \dedx spectrum where all the means and widths are constrained using enhanced pion, proton and electron samples.

\section{Results}

The transverse momentum spectra of charged pions, kaons and (anti)protons are shown in Fig.~\ref{Fig1} for the seven V0A multiplicity event classes measured in \ppb collisions. The low \pt results have been already published~\cite{Abelev:2013haa} and here the \pt reach is extended up to 15 GeV/$c$.
For \pt below 2 GeV/$c$ the spectra become harder as the multiplicity increases and the effect is stronger for heavier particles. This feature is well known from heavy nuclei collisions where it is attributed to the hydrodynamical evolution of the medium, and in fact the \pt spectra measured in high multiplicity \ppb collisions are better described by models which incorporate hydro~\cite{Abelev:2013haa}. However, it has been shown that also in \pp collisions simulated with Pythia 8 tune 4C~\cite{Corke:2010yf} the \pt spectra of identified particles as a function of multiplicity exhibit a qualitatively similar behaviour to that seen here. This behaviour is a consequence of the interactions among final partons coming from independent semi-hard scatterings which increases with increasing number of multi-parton interactions (MPI)~\cite{Ortiz:2013yxa}. 

\begin{figure}[htbp]
\begin{center}
   \includegraphics[width=0.9\textwidth]{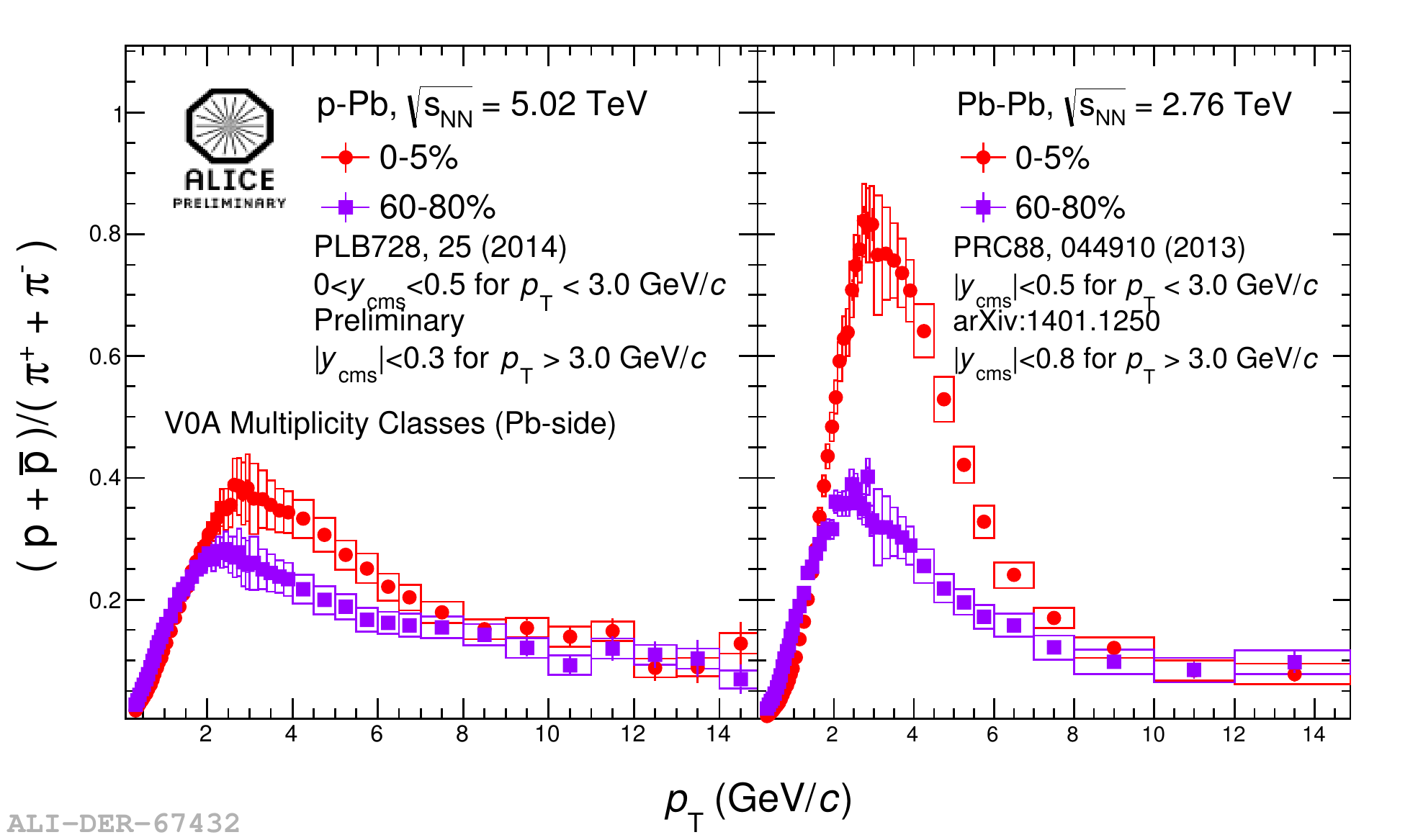}
   \caption{Proton-to-pion ratio as a function of \pt measured in \ppb (left) and \pbpb (right) collisions at $\sqrt{s_{\rm NN}}$ = 5.02 and 2.76 TeV, respectively. The systematic and statistical uncertainties are plotted as boxes and error bars, respectively.}
  \label{Fig3}
\end{center}
\end{figure}

One therefore cannot rule out alternative explanations, but interestingly, it illustrates that likely there is a strong coupling of this phenomenon to the underlying event also in \pp collisions. To study the evolution of the spectral shapes with multiplicity the blast-wave analysis has been performed and the results are shown in Fig.~\ref{Fig2}. This allows a comparison of the results from different colliding systems (\pp, \ppb and \pbpb) using a set of two parameters, which, in heavy ion collisions are typically connected with $T_{\rm kin}$ and $\langle \beta_{\rm T} \rangle$. Figure~\ref{Fig2} shows that a qualitatively similar behaviour is obtained for the three systems which were analysed, even in Pythia 8 events simulated with color reconnection.  
To study the effect on the \pt spectra directly, the proton-to-pion ratio was constructed, the results for \ppb and \pbpb collisions are presented in Fig.~\ref{Fig3} for two extreme multiplicity intervals. For \pt below (above) 2 GeV/$c$ the ratios exhibit a depletion (enhancement) going from low  to high multiplicity. The highest (lowest) multiplicity intervals give ratios which reach maxima at $\pt \approx 3$ GeV/$c$ amounting to $\approx$0.4 and $\approx 0.8$ ($\approx 0.28$ and $\approx$ 0.38) in \ppb and \pbpb collisions, respectively. Above 3 GeV/$c$, the ratios start to decrease down to $\approx0.1$ at \pt$\approx 10$ GeV/c, which according to~\cite{Abelev:2014laa} corresponds to the value measured for vacuum fragmentation (\pp collisions).

\section{Conclusions}

In this work the production of $\pi/{\rm K/p}$ measured in \ppb collisions have been presented up to $\pt = 15$ GeV/$c$. The baryon-to-meson ratios measured in \ppb and \pbpb collisions show an enhancement with respect to \pp collisions which reach their maxima at $\pt \approx 3$ GeV/$c$ and for higher \pt ($>10$ GeV/$c$) the ratios return to the value obtained for \pp collisions.  The evolution of the spectral shapes with the event multiplicity has a similar mass dependent systematics as observed in heavy nuclei collisions where it is associated to flow. Hydrodynamical calculations give the best description of the high multiplicity \pt spectra in \ppb and \pbpb collisions, but other mechanisms like MPI plus color reconnection also can produce flow-like patterns without the presence of any medium.





\bibliographystyle{elsarticle-num}

\bibliography{biblio}

\end{document}